\documentclass[11pt,prb,aps,showpacs,superscriptaddress]{revtex4}
\usepackage[latin9]{inputenc}
\usepackage{array}
\usepackage{amsmath}
\usepackage{amssymb}
\usepackage{esint}

\makeatletter

\DeclareRobustCommand{\greektext}{%
  \fontencoding{LGR}\selectfont\def\encodingdefault{LGR}}
\DeclareRobustCommand{\textgreek}[1]{\leavevmode{\greektext #1}}
\DeclareFontEncoding{LGR}{}{}
\DeclareTextSymbol{\~}{LGR}{126}
\newcommand{\lyxmathsym}[1]{\ifmmode\begingroup\def\b@ld{bold}
  \text{\ifx\math@version\b@ld\bfseries\fi#1}\endgroup\else#1\fi}

\providecommand{\tabularnewline}{\\}

\@ifundefined{textcolor}{}
{%
 \definecolor{BLACK}{gray}{0}
 \definecolor{WHITE}{gray}{1}
 \definecolor{RED}{rgb}{1,0,0}
 \definecolor{GREEN}{rgb}{0,1,0}
 \definecolor{BLUE}{rgb}{0,0,1}
 \definecolor{CYAN}{cmyk}{1,0,0,0}
 \definecolor{MAGENTA}{cmyk}{0,1,0,0}
 \definecolor{YELLOW}{cmyk}{0,0,1,0}
}

\usepackage{graphics}%\usepackage{epsf}
\usepackage{amsfonts}\usepackage{ulem}\usepackage{soul}

\makeatother

\begin{document}

\title{Magnetic properties and heat capacity of the three-dimensional frustrated
$S=1/2$ antiferromagnet PbCuTe$_{2}$O$_{6}$}

\author{B. Koteswararao}

\affiliation{Center of Condensed Matter Sciences, National Taiwan University,
Taipei 10617, Taiwan}

\affiliation{CeNSCMR, Department of Physics and Astronomy, and Institute of Applied
Physics, Seoul National University, Seoul 151-747, South Korea}

\author{R. Kumar}

\affiliation{Department of Physics, Indian Institute of Technology Bombay, Mumbai
400076, India}

\author{P. Khuntia}

\affiliation{Max Planck Institute for Chemical Physics of Solids, 01187 Dresden,
Germany}

\author{Sayantika Bhowal}

\affiliation{Department of Solid State Physics, Indian Association for the Cultivation
of Science, Jadavpur, Kolkata-700032, India}

\author{S. K. Panda}

\affiliation{Centre for Advanced Materials, Indian Association for the Cultivation
of Science, Jadavpur, Kolkata-700032, India}

\author{M. R. Rahman}

\affiliation{Center of Condensed Matter Sciences, National Taiwan University,
Taipei 10617, Taiwan}

\author{A. V. Mahajan}

\affiliation{Department of Physics, Indian Institute of Technology Bombay, Mumbai
400076, India}

\author{I. Dasgupta}

\affiliation{Department of Solid State Physics, Indian Association for the Cultivation
of Science, Jadavpur, Kolkata-700032, India}

\affiliation{Centre for Advanced Materials, Indian Association for the Cultivation
of Science, Jadavpur, Kolkata-700032, India}

\author{M. Baenitz}

\affiliation{Max Planck Institute for Chemical Physics of Solids, 01187 Dresden,
Germany}

\author{Kee Hoon Kim}

\email{khkim@phya.snu.ac.kr}

\affiliation{CeNSCMR, Department of Physics and Astronomy, and Institute of Applied
Physics, Seoul National University, Seoul 151-747, South Korea}

\author{F. C. Chou}

\email{fcchou@ntu.edu.tw}

\affiliation{Center of Condensed Matter Sciences, National Taiwan University,
Taipei 10617, Taiwan}
\begin{abstract}
We report magnetic susceptibility $(\chi)$ and heat capacity ($C_{p}$)
measurements along with \textit{ab-initio} electronic structure calculations
on PbCuTe$_{2}$O$_{6}$, a compound made up of a three dimensional
$3D$ network of corner-shared triangular units. The presence of antiferromagnetic
interactions is inferred from a Curie-Weiss temperature ($\theta_{CW})$
of about $-22$ K from the $\chi(T)$ data. The magnetic heat capacity
$C_{m}$ data show a broad maximum at $T^{max}\simeq1.15$ K {\normalsize{{$(i.e.$
}}}$T^{max}/\theta_{CW}\simeq0.05)$, which is analogous to the the
observed broad maximum in the $C_{m}/T$ data of a hyper-Kagome system,
Na$_{4}$Ir$_{3}$O$_{8}$. In addition, $C_{m}$ data exhibit a weak
kink at $T^{*}\simeq0.87$ K. While the $T^{max}$ is nearly unchanged,
the $T^{*}$ is systematically suppressed in an increasing magnetic
field ($H$) up to $80$ kOe. For $H\geq80$ kOe, the $C_{m}$ data
at low temperatures exhibit a characteristic power-law ($T$$^{\text{\textgreek{a}}}$)
behavior with an exponent $\lyxmathsym{\textgreek{a}}$ slightly less
than $2$. Hopping integrals obtained from the electronic structure
calculations show the presence of strongly frustrated 3$D$ spin interactions
along with non-negligible unfrustrated couplings. Our results suggest
that PbCuTe$_{2}$O$_{6}$ is a candidate material for realizing a
$3D$ quantum spin liquid state at high magnetic fields. 
\end{abstract}



\maketitle

\section{Introduction}

Interesting magnetic behavior in Heisenberg spin systems originates
from a network of some elementary motifs such as triangles or tetrahedra,
where the spins placed at their vertices interact with each other
via antiferromagnetic interactions. Due to the incompatibility of
the nearest neighbor interactions of a spin with respect to achieving
the lowest energy, these systems ideally do not have any conventional,
symmetry-breaking, phase transition. Instead, they often lead to exotic
ground states such as spin liquids and spin ice states.\cite{GFM2011,L. BalentsNature2010 ,J S Gardner RMP2010}
There are many well known experimental examples for the three dimensional
($3D$) networks of corner-shared tetrahedra, particularly in the
pyrochlore and spinel structures.\cite{J S Gardner RMP2010} On the
other hand, $3D$ networks of corner-shared triangles are relatively
less explored, despite the expectation that they would also display
novel ground states.

Two representative systems in the family of $3D$ corner-shared triangles
are Gd$_{3}$Ga$_{5}$O$_{12}$\cite{DG Onn PR 1967,S Hov JMMM1980,O A Petrenko2009-2010}
and Na$_{4}$Ir$_{3}$O$_{8}$.\cite{Y Okamoto Na4Ir3O8 PRL 2007}
The Gd$_{3}$Ga$_{5}$O$_{12}$ compound comprises of two inter-penetrating
hyper-Kagome networks, in which $3D$ coupled, corner-shared triangles
of Gd$^{3+}$ ($4f^{7}$ with $S=7/2$) spins are realized with coordination
number $z=4$ as in the case of the two dimensional $2D$ Kagome networks.
The Gd-system does not order down to $25$ mK in spite of a relatively
large Curie-Weiss temperature ($\theta_{CW})$ of $-2.3$ K,\cite{DG Onn PR 1967}
indicating that it is highly frustrated. It has a rich magnetic phase
diagram as investigated by magnetic susceptibility, heat capacity,\cite{S Hov JMMM1980}
and inelastic neutron scattering (INS) experiments.\cite{O A Petrenko2009-2010}
Na$_{4}$Ir$_{3}$O$_{8}$ forms a perfect hyper-Kagome network composed
of $5d$ Ir$^{4+}$ ions, and is considered as the first example of
a $3D$, $J_{eff}=1/2$ frustrated antiferromagnet with a quantum
spin liquid (QSL) ground state. The $5d$-based transition metal oxides
are known for having a large spin-orbit coupling and yet the Na$_{4}$Ir$_{3}$O$_{8}$
system retains the spin liquid ground state.\cite{G ChenPRL2008}
The exchange coupling between Ir$^{4+}$ spins is estimated to be
about $J/k_{B}\simeq-300$ K and the $\theta_{CW}$ is about $-650$
K while no signature of ordering is found down to $2$ K. Many theoretical
studies have also suggested that Na$_{4}$Ir$_{3}$O$_{8}$ is a promising
candidate for the $3D$ QSL ground state.\cite{MJLawlerPRL2008}

As far as the $3d$ transition metal based oxides are concerned, the
spin liquid state has been more widely found in the $2D$ corner-
and edge-shared frustrated triangular systems; $S=1/2$ Cu$^{2+}$-based
Kagome lattices ZnCu$_{3}$(OH)$_{6}$Cl$_{2}$ (Herberthsmithite\cite{J. S. Helton PRL2007}
and Kapellasite\cite{bfak prl2012}), $S=1/2$ V$^{4+}$-based $2D$
Kagome lattice {[}NH$_{4}${]}$_{2}${[}C$_{7}$H$_{14}$N{]}{[}V$_{7}$O$_{6}$F$_{18}${]}),\cite{L. ClarkPRL2013}
$S=1/2$ triangular lattices $\kappa$-(ET)$_{2}$Cu$_{2}$(CN)$_{3}$,\cite{Y. ShimizuPRL2003}
and Ba$_{3}$CuSb$_{2}$O$_{9}$.\cite{H. D. ZhouPRL2011,S Nakatsuji2012 }
To our knowledge, however, a candidate for the $S=1/2$, $3D$ QSL
ground state in the $3d$ transition metal oxides has not been reported
so far.

Herein, we introduce a new candidate with Cu-based $3D$ network,
PbCuTe$_{2}$O$_{6}$, which is isostructural to a mineral compound
Choloalite PbCuTe$_{2}$O$_{6}$Cl$_{0.333}$.\cite{D. W. Powell1994}
The compound crystallizes in a cubic structure with the space-group
$P4_{1}32$ (No. $213$) and has a $3D$ network of Cu atoms, where
the 1$^{st}$ nearest neighbors ($nn$) form uniform triangles, the
2$^{nd}$ $nn$ form a hyper-Kagome network and the 3$^{rd}$ $nn$
form uniform chains running in all the crystallographic directions.
The magnetic susceptibility $\chi(T)$ reveals a Curie-Weiss behavior
($\theta_{CW}\simeq-22$ K) down to $2$ K with no bifurcation between
zero-field-cooled (ZFC) and field-cooled (FC) curves at low fields.
However, the heat capacity $C_{p}(T)$ data at zero magnetic field
show a broad maximum at $T^{max}\simeq1.15$ K and also a weak kink,
at $T^{*}\simeq0.87$ K. These features in $C_{p}(T)$ data do not
appear to be a result of any conventional long-range order or spin
freezing. The weak kink at $T^{*}$ is systematically suppressed in
an increasing magnetic field ($H$) up to $80$ kOe. The $T^{max}$
is somewhat stable up to 80 kOe, but it is also slightly shifted and
suppressed for $H>80$ kOe. At $H\geq80$ kOe, the magnetic heat capacity
$C_{m}$ at low temperatures shows a power-law ($T$$^{\text{\textgreek{a}}}$)
behavior with an exponent $(\lyxmathsym{\textgreek{a}})$ slightly
less than $2$. The appearance of such a nearly $T^{2}$ behavior
in the $C_{m}(T)$ data under magnetic field infers the possible realization
of a\textbf{ }field tuned\textbf{ }QSL behavior in PbCuTe$_{2}$O$_{6}$.

\section{Experimental details}

The polycrystalline samples were prepared by solid-state reaction
method using high purity PbO, CuO, and TeO$_{2}$. The stoichiometric
amounts of starting materials were ground thoroughly, pelletized,
and subsequently fired at $500^{o}$C for 3 days, after sealing in
an evacuated quartz tube. To get single phase of PbCuTe$_{2}$O$_{6},$
the sample was subjected to three intermediate grindings at the same
temperature. The x-ray powder diffraction (XRD) pattern was recorded
at room temperature using PANalytical X'pert PRO diffractometer. The
crystal structure was refined by the Rietveld method on the powder
XRD data. The magnetic susceptibility of the sample was measured on
SQUID-VSM (Quantum design), in the temperature ($T$) range $2$ -
$300$ K, under a magnetic field $(H)$ of $10$ kOe. Heat capacity
$C_{P}$ measurements were performed down to $0.35$ K, using a PPMS\ with
$^{3}$He attachment, in fields up to $140$ kOe.

\section{Results}

\subsection{X-ray diffraction and structural features}

\begin{table}
\centering{}\caption{\label{tab:XYZ positions-1}Unit cell parameters of PbCuTe$_{2}$O$_{6}$
(space group: $P4_{1}32$ ($No.$ $213$) with the lattice constant
$a=12.49\pm0.005$ $\mathrm{\AA}$. }

\begin{tabular}{|l|l|l|l|l|l|}
\hline 
atom  & Wyckoff position  & $x$  & $y$  & $z$  & occ. \tabularnewline
\hline 
Te  & 24e  & 0.0812  & 0.4406  & 0.3401  & 1 \tabularnewline
\hline 
Pb1  & 8c  & 0.1918  & 0.1918  & 0.1918  & 1 \tabularnewline
\hline 
Pb2  & 4a  & 0.375  & 0.375  & 0.375  & 1 \tabularnewline
\hline 
Cu  & 12d  & 0.125  & 0.2321  & 0.4821  & 1 \tabularnewline
\hline 
O1  & 24e  & 0.0331  & 0.1205  & 0.2774  & 1 \tabularnewline
\hline 
O2  & 24e  & 0.1771  & 0.3164  & 0.3812  & 1 \tabularnewline
\hline 
O3  & 24e  & 0.1902  & 0.5240  & 0.2631  & 1 \tabularnewline
\hline 
\end{tabular}
\end{table}

In order to identify the crystal structure of PbCuTe$_{2}$O$_{6},$
initial parameters corresponding to the Choloalite mineral compound
PbCuTe$_{2}$O$_{6}$Cl$_{0.33}$\cite{D. W. Powell1994} were used.
The experimental powder diffraction pattern of PbCuTe$_{2}$O$_{6}$
matches well with the calculated powder pattern generated from the
mineral's data, using powder cell software,\cite{powdercel}. We then
refined the structure of PbCuTe$_{2}$O$_{6}$ by employing Rietveld
refinement software\cite{Carvajal 1993} (see Fig. 1). The refined
atomic coordinates are summarized in table \ref{tab:XYZ positions-1},
and the lattice constant was found to be $a\simeq12.49\pm0.005$ $\mathrm{\AA}$.
The residual factors are found to be $R_{p}\simeq3.099$ $\%$, $R_{wp}\simeq4.279$
$\%$, $R_{exp}\simeq2.024$ $\%$, $R_{Bragg}\simeq1.916$ $\%$,
and goodness of fit (GOF) $\simeq2.113,$ please modify this respectively.
To verify the quality of our fit done by Rietveld refinement method,
we also used Le Bail method\cite{LeBailMethod}, which is a simple
method to extract the integrated intensities for powder diffraction
pattern using only space group, lattice constants, profile, and back
ground parameters. Structural parameters are not used in the refinement
cycles. Using the same space group ($P4_{1}32$ ), it provided a reasonably
good fit to the experimentally observed data with a GOF about 3.213
and the obtained residual factors are also nearly consistent with
the Rietveld refinement.

The structure of PbCuTe$_{2}$O$_{6}$ consists of CuO$_{4}$ square
plaquettes, TeO$_{3}$ units, and Pb atoms, as shown in Fig. 2$(a)$.
As shown in Fig. 2$(b)$, the arrangement of Cu atoms in this structure
suggests it to be a $3D$ network of corner-shared triangles ($z=6$)
with a combination of $1^{st}$ nearest neighbor ($nn$) and $2^{nd}$
$nn$ interactions. The 1$^{st}$ $nn$ distance between Cu atoms
in each CuO$_{4}$ plaquette is about (4.37$\pm0.002$) $\mathrm{\AA}$
and an exchange coupling might be possible via O-Pb-O linkages with
a bond angle 91.3\textsuperscript{o} (see Fig. 2$(c)$). Taking this
bond length (4.37$\pm0.002$) $\mathrm{\AA}$ into consideration,
the Cu atoms appear to form isolated triangles in the structure, as
can also be seen in Fig. 2$(c)$ indicated by blue triangles. Herein,
we call this coupling $J_{tri}$. These blue triangles are connected
with each other via other equilateral triangles of side (5.60$\pm0.002$)
$\mathrm{\AA}$. Like $J_{tri}$ ($1^{st}$ $nn$), this $2^{nd}$
$nn$ path is clearly mediated by TeO$_{3}$ units (see Fig. 2$(a)$),
however it has a larger bond length than that of $1^{st}$ $nn$ blue-triangle.
The triangles with only this bond distance 5.60 $\mathrm{\AA}$ (red
color) form a network as in a hyper-Kagome lattice $i.e.,$ corner-shared
triangles with the exchange coupling $J_{hyper}$ (see Fig. 2$(d)$).
If we consider the $1^{st}$ $nn$ coupling to be very weak ($J_{tri}$$\simeq$0),
then the $2^{nd}$ $nn$ coupling leads to the formation of a hyper-Kagome
lattice exactly like that of Ir sites in Na$_{4}$Ir$_{3}$O$_{8}$
(which has $z=4$). But in reality, $J_{tri}$ might not be negligible
so PbCuTe$_{2}$O$_{6}$ has a higher connectivity than that of the
hyper-Kagome lattice. In addition to the 1$^{st}$ $nn$ and 2$^{nd}$
$nn$ couplings, we also looked at the $3^{rd}$ $nn$, which leads
to the formation of uniform chains with a bond length of 6.27 $\mathrm{\AA}$.
These uniform chains run along all the crystallographic axes (see
Fig. 2$(e)$). As the bond length is large, this chain coupling ($J_{chain}$)
is expected to be weaker than $J_{hyper}$.

\subsection{$\chi(T)$ and $C_{p}(T)$ results}

The magnetic susceptibility $\chi(T)$ $(=M(T)/H)$ is measured in
the temperature ($T$) range $2$- $300$ K in an applied field ($H$)
of $10$ kOe, as illustrated in Fig. 3$(a)$. The data follow a Curie-Weiss
behavior and no long-range order (LRO) is detected. The absence of
ZFC and FC bifurcation in thermal hysteresis measurements done with
$100$ Oe down to $2$ K (see inset of Fig. 3$(a)$), indicates the
absence of spin glass and/or freezing behavior. From a fit of the
$\chi(T)$ data to $\chi_{o}+\frac{C}{T-\theta_{CW}}$ in the $T-$range
from $20$ K to $300$ K, we get $\chi_{o}\simeq-(1.2\pm0.05)\times10^{-4}$
cm$^{3}$/mol Cu, $C=Ng^{2}\mu_{B}^{2}S(S+1)/3k_{B}\simeq(0.38\pm0.05)$
cm$^{3}$ K/mol Cu, and $\theta_{CW}\simeq-(22\pm0.5)$ K. Here $N_{A}$,
$g$, $\mu_{B}$, $k_{B}$, and $\theta_{CW}$ are Avogadro number,
Lande-$g$ factor, the Bohr magneton, the Boltzmann constant, and
the Curie-Weiss constant, respectively. The core diamagnetic susceptibility
($\chi_{dia}$) of PbCuTe$_{2}$O$_{6}$ was calculated to be $-1.65\times10^{-4}$
cm$^{3}$/mol formula unit from those of its ions (Pb$^{2+}$, Cu$^{2+}$,
and (TeO$_{3}$)$^{2-}$).\cite{P. W. Selwood 1956} The estimated
Van-Vleck paramagnetic susceptibility $\chi_{vv}(=\chi_{0}-\chi_{core})$
is found to be $\simeq4.5\times10^{-5}$ cm$^{3}$/mol, which is in
good agreement with that in the other cuprates.\cite{N. Motoyama 1996}
A negative $\theta_{CW}$ value suggests that spin correlations are
antiferromagnetic in nature. The estimated effective moment ($\mu_{eff}$)
of Cu$^{2+}$ is 1.73 $\mu_{B}$, same as the $S=1/2$ value.

In order to gain further insights into the ground state of the system,
the heat capacity ($C_{p}$) measurements were performed down to $350$
mK in magnetic fields up to 140 kOe. To extract the associated magnetic
heat capacity contribution $C_{m}$ in the absence of a suitable non-magnetic
analog of PbCuTe$_{2}$O$_{6}$, the lattice part was estimated by
fitting the raw data with an equation consisting of the linear combination
of one Debye and several Einstein terms (see the equation below).\cite{kittel}

\begin{flushleft}
{\tiny{{{{{{$C_{p}(T)=C_{D}\left[9k_{B}\left(\frac{T}{\theta_{D}}\right)^{3}\int\limits _{0}^{x_{D}}\frac{x^{4}e^{x}}{(e^{x}-1)^{2}}dx\right]+{\scriptstyle {\displaystyle \sum_{i}}}C_{E_{i}}\left[3R\left(\frac{\theta_{E_{i}}}{T}\right)^{2}\frac{exp\left(\frac{\theta_{E_{i}}}{T}\right)}{\left[exp\left(\frac{\theta_{E_{i}}}{T}\right)-1\right]^{2}}\right]$}}}}}}} 
\par\end{flushleft}

Here $R$ is the universal gas constant, $k_{B}$ is the Boltzmann
constant, $\theta_{D}$ and $\theta_{E_{i}}$ are the Debye and Einstein
temperatures. We were able to model lattice contribution with a combination
of one Debye term and three Einstein terms which fits the observed
data in the temperature range 30-120 K, as shown in Fig. 3(b). The
corresponding Debye and Einstein temperatures $\theta_{D}$ , $\theta_{E_{1}}$,
$\theta_{E_{2}}$, and $\theta_{E_{3}}$ were found to be 92 K and
79 K, 200 K, and 525 K, respectively and the associated coefficients
($C_{D}$, $C_{E_{1}}$, $C_{E_{2}}$,$C_{E_{3}}$) were fixed in
the ratio of 1:1:3:5, respectively (there are 10 atoms in the primitive
cell of PbCuTe$_{2}$O$_{6}$). The $C_{m}$ was obtained after subtracting
the estimated lattice contribution and the data are shown in the Fig.
4(a). We would like to note that we tried to fit the $C_{p}$ data
in various temperature ranges with a combination of Debye and Einstein
terms. Whereas the inferred $C_{m}$ below 3 K was found to be independent
of the fitting details and hence is considered reliable, the $C_{m}$
above 5 K was found to be very sensitive to the form used for the
lattice heat capacity.

The lattice heat capacity was found to be extremely weak at low $T$
and the estimated lattice contribution at $3$ K is approximately
$15$ times smaller than the total $C_{p}.$ As illustrated in the
inset of Fig. 3(b), the $C_{p}$ data at zero magnetic field show
an upturn below $2$ K and attain a broad maximum at $T^{max}\simeq1.15$
K. We could observe the broad maximum directly in our measured $C_{p}$
data due to the presence of negligible lattice contribution around
that temperature. The appearance of a broad maximum has also been
noticed in the plot of $C_{p}/T$ vs $T$ for many spin liquid systems.\cite{L. BalentsNature2010 ,Y Okamoto Na4Ir3O8 PRL 2007,J. G. Cheng PRL2011}
This suggests that the broad maximum in our $C_{p}(T)$ might be a
characteristic feature of highly frustrated spin systems. Generally,
the position of this broad maximum in $C_{p}$ data for frustrated
spin systems varies according to the strength of exchange couplings,
spin value and dimensionality of the system, and is found to roughly
appear at $T/\theta_{CW}=$ $0.02$ to $0.15$ (see Ref. \cite{Y Okamoto Na4Ir3O8 PRL 2007}).
In the titled compound, we found the broad maximum at $T/\theta_{CW}\simeq0.05$,
which is close to the value of 0.04 found in Na$_{4}$Ir$_{3}$O$_{8}$.
The calculated value of $C_{m}\theta_{CW}/T^{max}$ ($34$ J/mol K)
for PbCuTe$_{2}$O$_{6}$ is also similar to that of Na$_{4}$Ir$_{3}$O$_{8}$
indicating that the PbCuTe$_{2}$O$_{6}$ may have similar magnetic
correlations.

Further, a weak kink is also noticed below ($T^{max}$ ) at $0.87$
K in $C_{p}$ data, as marked by $T^{\ast}$ in the inset of Fig.
3(b). This anomaly at $T^{\ast}$ is also clearly visible as a small
hump in $C_{m}/T$ vs $T$ curve in Fig. 4(a). The origin for the
occurrence of this kink is not clear at this point. But, it does not
seem to be a standard LRO because of the following reasons: $(i)$
this weak-kink is not a $\lambda-$like anomaly and $(ii)$ the data
well below the kink do not follow exactly a $T^{3}$ behavior (antiferromagnetic
magnon dispersion). The absence of LRO down to $350$ mK with a relatively
large $\theta_{CW}$ of $-22$ K and the appearance of a broad maximum
at $1.15$ K in $C_{p}$ data all together provide a hunch of possible
quantum spin liquid (QSL) ground state.

\subsection{$C_{m}(T)$ in magnetic fields}

The plot of magnetic heat capacity $C_{m}$ divided by $T$ ($C_{m}$/$T$)
as a function of $T$ in magnetic fields upto 140 kOe is shown in
Fig. 4. As it is seen in the Fig. 4, the $T^{max}$ is stable upto
80 kOe and it is slightly suppressed for fields $H\geq80$ kOe. On
the other hand, the weak-kink at $T^{\ast}$ is suppressed completely
by a magnetic field of 80 kOe. The observed $T^{*}$ and $T^{max}$
(chosen from the plots of $C_{m}/T$ vs $T$ and $C_{m}/T^{2}$ vs
$T$) are plotted in Fig. 5.

To explore the nature of excitations at low $T$, we fit the $C_{m}$
data with a power-law ($C_{m}\propto T^{\alpha}$). As it is evident
from the Fig. 4(a), the zero field $C_{m}/T$ data do not follow a
single power law below the $T^{max}$ due to the presence of $T^{*}$.
Upon increasing the magnetic field, the $T^{*}$ is systematically
suppressed and is apparently reduced to zero by magnetic fields greater
than 80 kOe. Moreover, the $C_{m}/T$ data seem to follow a single
power law behavior for $H\ge80$ kOe. We then fitted the data with
a single slope ($T^{\lyxmathsym{\textgreek{a}}}$) behavior in the
temperature range $0.35-0.6$ K below $T^{max}$ (see Fig. 4(e-h)).
The obtained values are near to $1.9\pm0.1$, slightly less than 2.
The appearance of a near $T^{2}$ behavior in $C_{m}$ data has been
observed for other reported $3D$ QSL systems Na$_{4}$Ir$_{3}$O$_{8}$
and cubic-Ba$_{3}$NiSb$_{2}$O$_{9}$.\cite{Y Okamoto Na4Ir3O8 PRL 2007,J. G. Cheng PRL2011}
In case of Na$_{4}$Ir$_{3}$O$_{8}$, the value of $\alpha$ is found
to be in between 2 and 3, while for cubic-Ba$_{3}$NiSb$_{2}$O$_{9}$
the observed $\alpha$ value was $2.0\pm0.2$. On the other hand,
in our case, the observed $\alpha$ value deviates slightly compared
to the above examples. The reason might be due to having a different
$3D$ frustrated spin structure for PbCuTe$_{2}$O$_{6}$. Several
theories also predict that the $T^{2}$ dependency of the magnetic
heat capacity is expected for the $3D$ QSLs.\cite{MJLawlerPRL2008}
We note that the variation of $\alpha$ or $C_{m}$ in PbCuTe$_{2}$O$_{6}$
at higher magnetic fields is apparently due to the comparable strengths
of Zeeman energies and the super-exchange couplings in this system.

\subsection{First principles electronic structure calculations }

\begin{table}
\begin{centering}
\caption{\label{Hoppings-1}The details of bond length, hopping integrals and
the relative exchange couplings between the Cu atoms. }

\par\end{centering}

\centering{}%
\begin{tabular}{|>{\raggedright}p{0.5cm}|>{\raggedright}p{0.8cm}|>{\raggedright}p{1.5cm}|>{\raggedright}p{1.7cm}|>{\raggedright}p{1.6cm}|}
\hline 
\emph{$t_{i}$}  & $J_{i}$  & Cu-Cu distance

($\mathrm{\AA}$)  & Hopping parameters (meV)  & $\frac{J_{i}}{J_{hyper}}=\left(\frac{t_{i}}{t_{2}}\right)^{2}$ \tabularnewline
\hline 
$t_{1}$  & $J_{tri}$  & 4.37  & 45.59  & 0.54 \tabularnewline
\hline 
$t_{2}$  & $J_{hyper}$  & 5.60  & 62.18  & 1.00 \tabularnewline
\hline 
$t_{3}$  & $J_{chain}$  & 6.27  & 54.47  & 0.77 \tabularnewline
\hline 
\end{tabular}
\end{table}

In order to understand the electronic structure of PbCuTe$_{2}$O$_{6}$
and to identify the dominant exchange paths, we have performed first-principles
density functional theory (DFT) calculations within the local density
approximation (LDA) using projector augmented-wave (PAW) method\cite{P. E. Blochl1994,g kresse1999}
encoded in the Vienna $ab$ $initio$ simulation package (VASP).\cite{g kresse1993,g kresse1996}
The energy cut-off for the plane wave expansion of the PAW's was taken
to be $550$ eV and a (8 $\times$ 8 $\times$ 8) $k$-mesh has been
used for self consistency. The non-spin-polarized band structure for
PbCuTe$_{2}$O$_{6}$ is displayed in Fig. 6. The bands are plotted
along the various high symmetry points of the Brillouin zone corresponding
to the cubic lattice. All the energies are measured with respect to
the Fermi level of the compound. The characteristic feature of the
non-spin-polarized band structure, displayed in Fig. 6(a), is the
isolated manifold of twelve bands near the Fermi level $(E_{F})$.
These bands are predominantly of Cu $d_{x^{2}-y^{2}}$ character in
the local frame of reference, arising from twelve Cu atoms in the
unit cell, where Cu is at the square planar environment of O ions.
The crystal field splitting corresponding to a square planar environment
is shown in the inset of Fig. 6(a). Since Cu is in 2+ state $i.e,$
in the $d^{9}$ configuration, these isolated bands are half filled
and are separated from the other Cu $d$ bands by a gap of $0.9$
eV. These set of twelve bands are responsible for the low energy physics
of the material.

To extract a low energy model Hamiltonian, we have constructed the
Wannier functions for these bands using the VASP2WANNIER and the WANNIER90
codes.\cite{aa mostofi2008} The Wannier-interpolated bands along
with the LDA bands are displayed in Fig. 6(b) and the agreement is
quite remarkable.\textbf{ }The various hopping interactions ($t_{n}$)\textbf{
}obtained in this method are listed in Table \ref{Hoppings-1}. The
super-exchange interactions between the Cu atoms can be estimated
using the relation $J=\frac{4t_{n}^{2}}{U_{eff}}$, where $U_{eff}$
is the effective onsite Coulomb interaction and $t_{n}$ corresponds
to the hopping via super-exchange paths. The relative exchange couplings
are also displayed in Table \ref{Hoppings-1}. From the Table \ref{Hoppings-1},
the $J_{hyper}$ is found to be the strongest interaction in the system,
while the $J_{tri}$ and $J_{chain}$ are also substantial. In order
to understand why $J_{hyper}$ is the most significant interaction
and what are the super-exchange paths, we look at the interesting
crystal geometry of this system in\textbf{ }Fig. 2\textbf{. }The $2^{nd}$
$nn$ Cu atoms interact via Cu-O-Te-O-Cu path as shown in\textbf{
}Fig.\textbf{ }2$(d)$. A strong hybridization exists between the
Cu $d_{x^{2}-y^{2}}$ with O $p_{x}$ and O $p_{y}$ orbitals via
$\sigma$-bonding and also with cation Te$^{4+}$ ($5s^{2}$). Moreover,
this coupling involves shorter Te-O bonds (2.01 $\mathrm{\AA}$),
whereas other couplings, $J_{tri}$ and $J_{chain}$, involve a large
Pb-O bond distance 2.58 $\mathrm{\AA}$. As a consequence, the $t_{2}$
hopping becomes most significant and gives rise to a hyper-Kagome
network with the coupling $J_{hyper}$. This is also corroborated
by a plot of the Wannier function for the Cu $d_{x^{2}-y^{2}}$ orbital
(See Fig. \textbf{7}). The plot of Wannier function illustrates the
importance of Te atoms in the exchange path $J_{hyper}$ and it also
reveals that Cu $d_{x^{2}-y^{2}}$ orbital forms strong $pd\lyxmathsym{\textgreek{sv}}$
antibonding with the neighboring O $p_{x}$ and O $p_{y}$ orbitals.
From the tail of the Wannier function, we find that the Te also hybridizes
strongly with Cu $d$ and O $p$ orbitals and this hybridization is
responsible for the strong $2^{nd}$ $nn$ interaction between the
Cu atoms. The next dominant interaction is the one between the $3^{rd}$
$nn$ Cu atoms $(J_{chain})$. As can be seen in Fig. $2(e)$, the
Cu atoms follow the path Cu-O-Pb-O-Cu and thereby form a $1D$ chain,
which is actually a non-frustrated interaction. For this coupling,
the Cu $d_{x^{2}-y^{2}}$ strongly hybridizes with the O $p_{x}$
orbital via $\sigma$-bonding and also weakly via Pb atoms. The bond
angle of O-Pb-O is 154$^{o}$ along this path. The $3^{rd}$ strongest
exchange path is $J_{tri}$ between the $1^{st}$ $nn$ Cu atoms.
Here the exchange interactions between the Cu atoms follow the Cu-O-Pb-O-Cu
path with a bond angle of O-Pb-O to be 92$^{o}$. Due to this relatively
small bond angle of $O-$Pb-O, the $J_{tri}$ coupling is apparently
weaker than $J_{hyper}$. In the light of LDA results, we can conclude
that PbCuTe$_{2}$O$_{6}$ indeed has a hyper-Kagome network, as seen
in Na$_{4}$Ir$_{3}$O$_{8}$, but with the considerable residual
frustrated $(J_{tri})$ and non-frustrated $(J_{chain})$ couplings,
respectively.

\section{discussion}

While $3D$ spin systems exhibit mostly ordered behavior in the absence
of a special geometry, this is prevented in $1D$ or $2D$ spin systems
due to the underlying strong quantum fluctuations. In some cases,
the $3D$ spin systems can also avoid this spin-solid state and stay
in the disordered and/or spin liquid ground state due to the strong
frustration offered by a combined effect of geometry and the low value
of spin. Such a $3D$ spin liquid ground state originating from strong
frustration has been so far realized only in a hyper-Kagome lattice
Na$_{4}$Ir$_{3}$O$_{8}$. In this sense, PbCuTe$_{2}$O$_{6}$ can
be a second candidate system to expect such a ground state since it
has a similar space group symmetry with Na$_{4}$Ir$_{3}$O$_{8}$
and its spins form a corner-shared $3D$ frustrated network. On the
other hand, the spin interaction in PbCuTe$_{2}$O$_{6}$ is a bit
different from that of Na$_{4}$Ir$_{3}$O$_{8}$ in the sense that
it has two comparable $J_{hyper}$ and $J_{tri}$ with z=6 connectivity
and additional non-frustrated ($J_{chain}$) coupling. Even though
the details are somewhat different, PbCuTe$_{2}$O$_{6}$ still exhibit
clearly the several evidences of strong spin frustration: the absence
of LRO down to $350$ mK, the appearance of a broad maximum in $C_{p}$
at $1.15$ K, and a characteristic power-law behavior with an exponent
close to 2 at low $T$ for magnetic fields above 80 kOe. These experimental
features seem to be rather similar to those observed in Na$_{4}$Ir$_{3}$O$_{8}$.

In contrast, the presence of a weak kink at $T$$^{*}$ of $C_{p}$
data is a distinct feature that was not observed in Na$_{4}$Ir$_{3}$O$_{8}$.
In order to understand the true origin of $T$$^{*}$, here we discuss
two scenarios based on the existing results in other frustrated spin
systems. The $1^{st}$ scenario is that it might be associated with
a thermodynamic phase transition different from the standard LRO,
with significant short-range correlations. A similar low $T$ kink
was also found in $C_{p}/T$ vs. $T$ data in a polycrystalline Volborthite
Cu$_{3}$V$_{2}$O$_{7}$.2H$_{2}$O.\cite{S YamashitaJPSJ2010} Using
$^{51}$V NMR, $\mu$SR, and inelastic neutron scattering (INS) measurements,
it turned out to be a kind of thermodynamic transition with dense
low-energy excitations and persistent spin dynamics down to $20$
mK.\cite{F Bert2005} Owing to the presence of additional non-frustrated
exchange coupling $J_{chain}$, PbCuTe$_{2}$O$_{6}$ might be also
prone to have an unusual thermodynamic transition at a finite $T$.
% We suspect that $J_{chain}$ might play a significant role in having the kink at $T^{\ast}$ in $C_{p}$. 
A similar scenario exists in $S=1/2,$ $3D$ frustrated double-pervoskite
Ba$_{2}$YMoO$_{6}$, in which a valence-bond-glass ground state was
proposed due to the involvement of the $2^{nd}$ $nn$ coupling.\cite{MAdeVries2010}

The $2^{nd}$ possibility as the origin of $T$$^{*}$ could be a
signature of a valence bond crystal (VBC), which preserves the rotational
symmetry but breaks the translational invariance. Recently, Bergholtz
\textit{et al}. (Ref. 34) have theoretically proposed a valence bond
crystal (VBC) ground state for the $S=1/2$ Heisenberg model in the
hyper-Kagome lattice, a contender to the other QSL proposals.\cite{MJLawlerPRL2008}
In their proposal, the lack of magnetic ordering as well as the existence
of a peak in $C_{m}/T$ have been identified as the probable tests
for the VBC picture. Since our system exhibits also a weak-kink in
the heat capacity, this possibility might be worth consideration.
On the other hand, the higher connectivity $(z=6)$ of PbCuTe$_{2}$O$_{6}$
than that of hyper-Kagome $(z=4)$ may have to be distinguished before
considering the application of the theoretical proposal. Exploring
the true nature of $T$$^{*}$ in PbCuTe$_{2}$O$_{6}$ is thus likely
to provide an opportunity to understand the unusual ordering in highly
frustrated spin networks.

\section{summary}

In summary, the structure of PbCuTe$_{2}$O$_{6}$ is made up of a
$3D$ network with various frustrated ($J_{hyper}$ \& $J_{tri}$)
and non-frustrated exchange couplings ($J_{chain})$. The $\chi(T)$
data show relatively large AF correlations and the absence of any
conventional LRO down to $350$ mK. The $C_{m}$ data in zero field
evidence a broad maximum at $T{}_{max}$$/$$\theta_{CW}\simeq$$0.05$
and $\frac{C{}_{m}\theta_{CW}}{T}\simeq34$ J/mol K, which are similar
to the values found in another $3D$ QSL system Na$_{4}$Ir$_{3}$O$_{8}$.
A weak kink found at $0.87$ K in the $C_{m}$ data is suppressed
completely by a magnetic field of $80$ kOe and then the $C_{m}$
data follow nearly a $T^{2}$ behavior at lower temperatures. From
our experimental observations, we suggest that PbCuTe$_{2}$O$_{6}$
might be an example of a field-induced QSL\textbf{ }in the family
of $3D$ frustrated Cu-based systems. To unveil more about the spin
excitations of the ground state and field-induced states in detail,
one needs to further explore this system intensively including NMR,
$\mu SR$, and inelastic neutron scattering measurements at various
energy scales.

The authors BK and FCC acknowledge the support from the National Science
Council of Taiwan under project number NSC-102-2119-M-002-004. AVM
thanks the Department of Science and Technology, Government of India
for financial support. R. Kumar and S. Bhowal acknowledge CSIR India
for financial support. The work at SNU was supported by the National
Creative Research Initiative (2010-0018300). We thank Fabrice Bert
for fruitful discussions.

\end{document}